\documentclass[preprint,12pt]{elsarticle}
\usepackage[utf8]{inputenc}

\usepackage{amssymb}
\usepackage{amsmath}
\usepackage{siunitx}

\usepackage{graphicx}

\usepackage{booktabs}

\journal{Nuclear Instruments and Methods in Physics Research Section A}

\begin{document}

\begin{frontmatter}



\title{Simulation of irradiated hybrid planar pixels modules at fluences expected at HL-LHC}


\author{M. Bomben} %% Author name

\affiliation{organization={APC, Universite Paris Cite, CNRS/IN2P3},%Department and Organization
            addressline={10 rue Alice Domon et Leonie Duquet}, 
            city={Paris},
            postcode={75013}, 
            %state={},
            country={France}}

%% Abstract
\begin{abstract}

Signal loss is the main limitation on tracking/vertexing performance due to radiation damage effect to hybrid pixel detectors when irradiated at fluences expected at High Luminosity LHC (HL-LHC).
It is important to have reliable predictions on the charge collection performance after irradiation in order to predict operational voltage values and test tracking algorithms robustness.
In this paper  the validation of combined TCAD and Monte Carlo simulations of hybrid silicon planar pixels sensors will be presented. 
In particular different trapping models will be compared to identify the one giving the best predictions.
Eventually predictions on the collected charge performance of planar pixels modules at HL-LHC will be discussed.
\end{abstract}



%% Keywords

\end{frontmatter}



\section{Introduction}
\label{sec:intro}

Silicon radiation detectors are exposed to unprecedented hadron fluences at Large Hadron Collider (LHC) experiments ATLAS, CMS and LHCb~\cite{Dawson:2764325}.  
Deep defects are created in the Si bandgap by hadrons determining modifications to the space charge distribution and to the generation rate~\cite{Moll2018}. As a result the leakage current and the  
depletion voltage of the detector increases, while carriers being trapped result in a reduction of the signal amplitude. Charge trapping can lead to induced signals being below detection threshold, 
determining the distortion of cluster shapes, resulting in degradation of spatial resolution and hit efficiency~\cite{ATL-PHYS-PUB-2022-033}. 
It is extremely important to have Monte Carlo (MC) simulated events that can reproduce the loss of collected signal with the increase of integrated luminosity, hence fluence. 
The ATLAS~\cite{RadDamagePaper2019} and CMS~\cite{Swartz:2008oU} collaboration have developed and implemented algorithms that reproduce with percent precisioni~\cite{EPS2023Bomben} 
the loss of collected charge with the accumulated fluence. 

The LHC will be upgraded into a high luminosity collider (HL-LHC), capable of delivering proton-propton collisions at a rate five to seven times larger than today with the goal to accumulate 
a data set ten times larger than the actual one in about ten years of operations~\cite{HLLHC}. Such an increase of collisions rate and particles poses stringent constraints on the silicon detectors in terms of 
radiation damage - up to 10 times larger than today - to the point that a new inner detector is needed in both experiments. ATLAS will implement a new all silicon inner detector, the Inner 
Tracker (ITk)~\cite{ATLASITkPixelTDR,ITkStripsTDR}. Hybrid silicon n-on-p pixel modules will be used in the core part of ITk, with strip detectors at larger radii. 
The innermost part of the ITk Pixel detector will be instrumented with 3D sensors will all the rest will be equipped with planar ones; more details on the ITk layout can be found in~\cite{itk_tracking_perf_2024}. 
The largest fluence to be integrated by 3D and planar sensors is of about $1.6\times10^{16}$ and $3.5\times10^{15}$~\SI{}{n_{eq}/cm^2} respectively.

It is extremely important to have predictions on charge collection performance after such large fluences and the combination of TCAD~\footnote{Technology Computer Aided Design} and MC tools like 
Allpix$^2$~\cite{Allpix2} is perfect as with the first a precise simulation of the electric field inside the sensor is possible while the latter is adapted for simulating particles impinging on the detector under different 
conditions (temperature, threshold, angle, etc.). This approach is being used for example in several cases in detector development for high energy physics~\cite{WENNLOF2025170227,s24123976}.

At the moment no data from irradiated modules equipped with the final version of readout chip (ITkPixV2~\cite{pottier2025testinglimitsitkpixv2atlas}) for the ATLAS ITk pixel detector exist. 
In order to assess the expected collected charge existing data on similar devices after HL-LHC like fluences have been identified; in particular the CMS tracker group 
investigated passive CMOS sensors~\cite{CMS_passive_CMOS_unirradiated,CMS_irradiated_Pisa_passive_cmos} 
bump bonded to an RD53A readout chip~\cite{Garcia-Sciveres:2287593}. Results from 
testbeams for n-on-p sensors with a thickness of 150~\SI{}{\micro\meter} were reported. These results have been used to validate the predictions of TCAD and Allpix$^2$.

In Section~\ref{sec:setup} the simulation setup will be presented, then its validation will be discussed~(Section~\ref{sec:validation}). The expected charge collection performance of ATLAS ITk planar pixel modules 
will be presented in Section~\ref{sec:performance}; Section~\ref{sec:conclusions} will conclude the paper. 

\section{Simulation Setup}
\label{sec:setup}

The simulation of the irradiated silicon pixels sensors has been carried out using the Geant4~\cite{Geant41,Geant42,Geant43} based Allpix$^2$ (v3.2.0) MC simulation framework, with  precise electric field 
and weighting potential~\cite{ShockleyPot,Ramo} maps calculated using Silvaco TCAD\footnote{https://silvaco.com/tcad/} tools as input. 
The simulated sensor was  a 150~\SI{}{\micro\meter} thick n-on-p pixels with a 50$\times$50~\SI{}{\micro\meter}$^2$ pitch. Details of the setup of TCAD and Allpix$^2$ simulations 
will be presented in the following.

\subsection{TCAD setup}
\label{sec:TCADsetup}
Silvaco TCAD tools were used to perform device simulation with the goal of obtaining a 3D map of the weighting potential and of the electric field, the latter  at different voltages after several irradiation fluences. 
The TCAD simulation setup included  Fermi-Dirac statistics, Shockley-Read-Hall recombination, 
trap assisted tunneling~\cite{HurkxTAT}, mobility model~\cite{CANALI19751122}, bandgap narrowing~\cite{KLAASSEN1992125} and impact ionization~\cite{VANOVERSTRAETEN1970583}. 
Given the range of fluences of interest the radiation damage model~\cite{FOLKESTAD201794} developed 
for the upgrade of the LHCb Velo detector~\cite{LHCbVeloUpgradeTDR} has been chosen; this model features two deep acceptor and one donor state. 

\begin{figure}[!htb]
   \centering
   \includegraphics[width=0.49\textwidth]{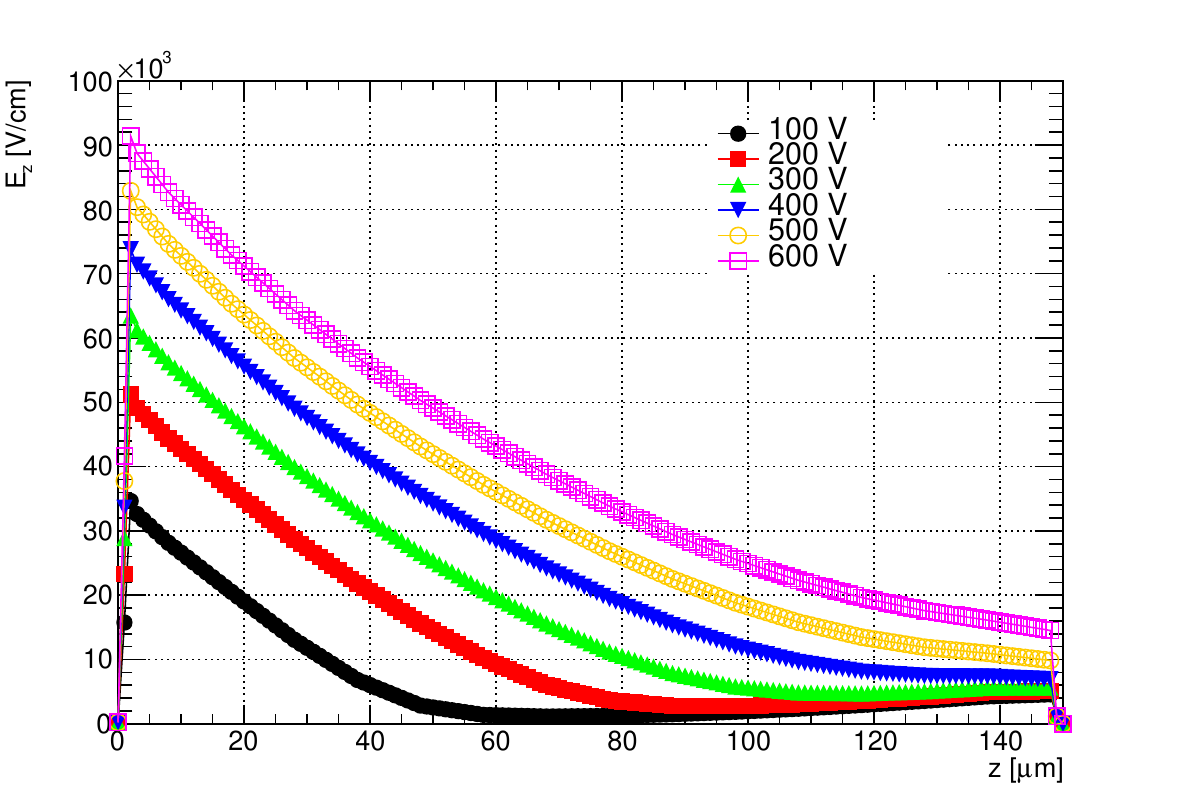} 
    \includegraphics[width=0.49\textwidth]{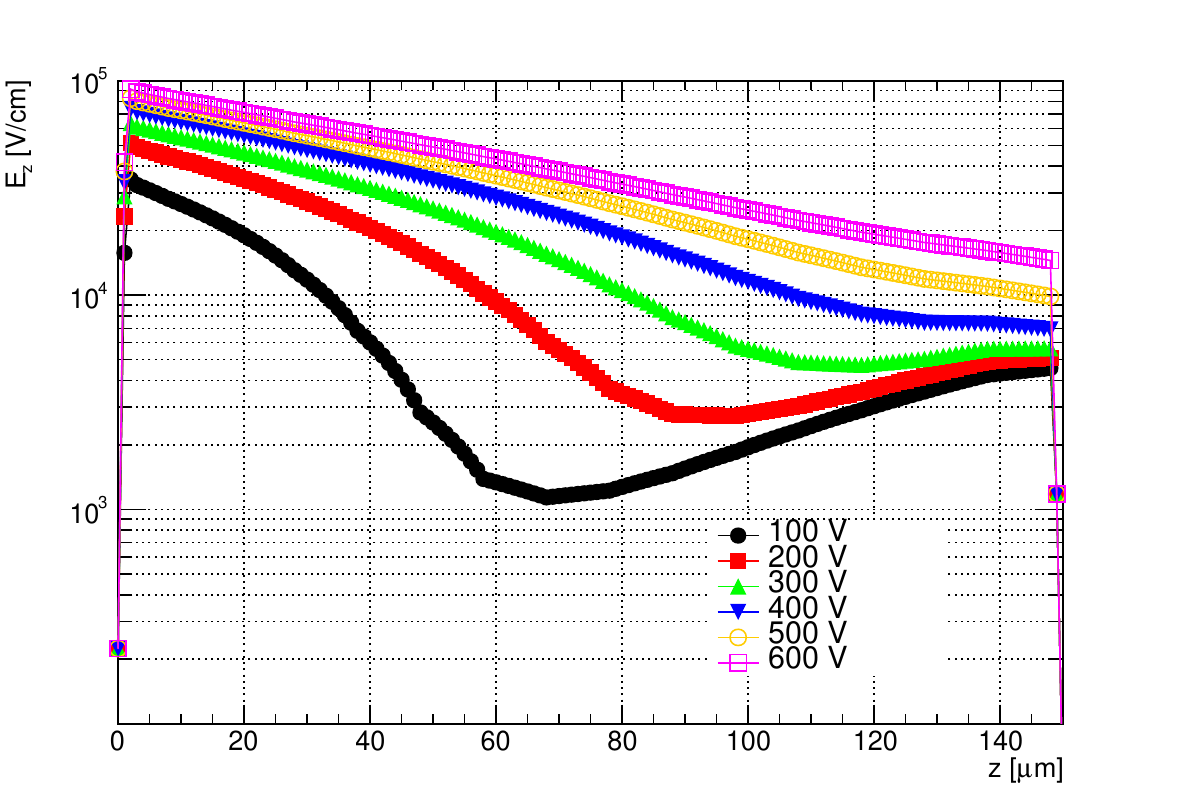} 
   \caption{Component along the sensor bulk of the electric field as a function of the bulk depth after a simulated fluence of $2.1\times10^{15}$~\SI{}{n_{eq}/cm^2} at different voltages. 
   (left) linear scale; (right) log scale. The junction side is at $z = 0$.}
   \label{fig:Ez_vs_z}
\end{figure}

In Figure~\ref{fig:Ez_vs_z} the simulated electric field component along the bulk is reported as a function of the depth in the bulk for different bias voltages when the simulated fluence was 
$2.1\times10^{15}$~\SI{}{n_{eq}/cm^2}; the field was evaluated in the center of a single pixel cell.

As it can be seen the electric field profile is  non linear at moderate voltages and  is characterised by a large value at the junction side ($z = 0$).  At 100~V there is a deep minimum close to the mid-plane of 
the sensor; as voltage increases the minimum moves to the backside. At 300~V the field is in excess of 5~kV/cm everywhere 
in the bulk. At 400~V the minimum is no longer there; on the contrary 
a sort of plateau of about 7~kV/cm with a breadth of approximatively \SI{30}{\micro\meter}  appears at the backside.

\subsection{Allpix$^2$ setup}
\label{sec:allpixsetup}

The configuration of Allpix$^2$ simulations included choice of temperature, mobility model, noise and threshold of the digitization step, the trapping model and the type of beam. 
In Table~\ref{tab:configuration} the 
complete list of models and related parameters values is reported.

% Requires the booktabs if the memoir class is not being used
\begin{table}[!htb]
   \centering
   %\topcaption{Table captions are better up top} % requires the topcapt package
   \begin{tabular}{@{} lcr @{}} % Column formatting, @{} suppresses leading/trailing space
      \toprule
     % \multicolumn{2}{c}{Item} \\
     % \cmidrule(r){1-2} % Partial rule. (r) trims the line a little bit on the right; (l) & (lr) also possible
           & Parameter & Value/Model \\
      \midrule
       Number of events    &  - & 5000 \\
       \hline
       Beam & & \\          
            & particle       & $\pi^+$ \\
             & energy  & 120~GeV \\
             & angle & normal \\
             \hline
     Sensor       &   &  \\
       & temperature   &  -20$^\circ$~C  \\
       & mobility & Canali~\cite{CANALI19751122} \\
       & trapping & See~Sect.~\ref{sec:trapping} \\
       \hline
     Digitization & & \\
                        & threshold & 600~e \\
                        & threshold dispersion & 115~e \\
                        & noise & 75~e \\  
      \bottomrule
   \end{tabular}
   \caption{Configuration of Allpix$^2$ simulations.}
   \label{tab:configuration}
\end{table}

Concerning the trapping model the choice was made after testing several of them against data. This will be explained in detail in Section~\ref{sec:validation} while  the different 
tested trapping models are presented in the following.

\subsubsection{Trapping models}
\label{sec:trapping}

Trapping rates $\tau_{e,h}^{-1}$ of electrons and holes in irradiated silicon bulk increase with the integrated fluence $\Phi$. In the context of this work four different trapping models have been tested: CMS (short for CMS tracker)~\cite{Allpix2}, 
Ljubljana~\cite{Trapping}, Atlas~\cite{RadDamagePaper2019} and Mandi\'{c}~\cite{Mandic2020}. A brief recap of the models is provided below, together with expected trapping times at two target fluences, 
$\Phi = 2.1\times10^{15}$~and~$1\times10^{16}$~\SI{}{n_{eq}/cm^2}; 
these fluences have been selected among those for which sample measurements were reported in~\cite{CMS_irradiated_Pisa_passive_cmos}.

\paragraph{\bf CMS}
The CMS model is based on the interpolation of results on trapping rates reported in~\cite{Adam_2016} where p-on-n diodes were measured after fluences up to $\Phi = 3\times10^{15}$~\SI{}{n_{eq}/cm^2}. 
The resulting 
expression for the trapping rates is:
\begin{equation}
\tau_{e,h}^{-1} = \beta_{e,h}\cdot\Phi + \tau_{0_{e,h}}^{-1}
\label{eq:CMS}
\end{equation}
where $\beta_{e,h} = 1.7$~and~2.8$\times 10^{16}\;$\SI{}{cm^{2}~ns^{-1}} and $ \tau_{0_{e,h}}^{-1} = 0.11$~and~$0.09$~ns$^{-1}$ for electrons and holes, respectively.

\paragraph{\bf Ljubljana}
The Ljubljana model is based on p-on-n diodes irradiated up to $\Phi = 2.4\times10^{14}$~\SI{}{n_{eq}/cm^2}. The resulting 
expression for the trapping rates is:
\begin{equation}
\tau_{e,h}^{-1} = \beta_{e,h}(T)\cdot\Phi
\label{eq:Ljubljana}
\end{equation}
where $\beta_{e,h}(T) = \beta_{e,h}(T_0)\left(\frac{T}{T_0}\right)^\kappa$~and~$T_0 = -10^{\circ}C$. 
The following values are used: $\beta_{e,h}(T_0) = 5.6$~and~7.7~$\times 10^{16}\;$\SI{}{cm^{2}~ns^{-1}} and $ \kappa_{e,h} = -0.86$~and~$ - 1.52$ for electrons and holes, respectively.

\paragraph{\bf Atlas}
The Atlas model was developed based on existing literature in preparation of the radiation damage digitizer for the Atlas pixel detector~\cite{RadDamagePaper2019}. 
The expression for the trapping rate is:
\begin{equation}
\tau_{e,h}^{-1} = \beta_{e,h}\cdot\Phi 
\label{eq:Atlas}
\end{equation}
where $\beta_{e,h} = 4.5$~and~6.5$\times 10^{16}\;$\SI{}{cm^{2}~ns^{-1}}  for electrons and holes, respectively.

\paragraph{\bf Mandi\'c}
The Mandi\'c\footnote{At the time of the article the implementation in Allpix$^2$ of the Mandi\'c trapping model differs from the one published. The model was reimplemented correctly by the author.} model~\cite{Mandic2020,Mandic2021} was developed for a range of very large fluences, going from  $\Phi = 5\times10^{15}$ to $1\times10^{17}$~\SI{}{n_{eq}/cm^2}.
The expression for the trapping time is:
\begin{equation}
\tau = c\cdot\left(\frac{\Phi}{1\times10^{16}}\right)^{\kappa}
\label{eq:Mandic}
\end{equation}
The following values are used: $c = 0.54$~ns and $\kappa =  -0.62$.

In Table~\ref{tab:trappingtimes} trapping times for the four models for two target fluences $\Phi = 2.1\times10^{15}$~and~$1\times10^{16}$~\SI{}{n_{eq}/cm^2}.

% Requires the booktabs if the memoir class is not being used
\begin{table}[!htb]
   \centering
   %\topcaption{Table captions are better up top} % requires the topcapt package
   \begin{tabular}{@{} ccccc @{}} % Column formatting, @{} suppresses leading/trailing space
      \toprule
     % \cmidrule(r){1-2} % Partial rule. (r) trims the line a little bit on the right; (l) & (lr) also possible
       & CMS & Atlas & Ljubljana & Mandi\'c \\
      \midrule
      \multicolumn{5}{c}{$\Phi = 2.1\times10^{15}$~\SI{}{n_{eq}/cm^2}} \\
          electrons     & 2.20 & 1.06 & 0.82 & 1.42\\
           holes    & 1.47 & 0.73 & 0.80 & 1.42 \\ 
     \hline
      \multicolumn{5}{c}{$\Phi = 1\times10^{16}$~\SI{}{n_{eq}/cm^2}} \\
          electrons     & 0.55 & 0.22 & 0.17 & 0.54 \\
           holes    & 0.35 & 0.16 & 0.17 & 0.54 \\ 
      \bottomrule
   \end{tabular}
   \caption{ \label{tab:trappingtimes}Trapping times (in ns) for two target fluences for the four  investigated trapping models.}
  \end{table} 

Given the order of magnitude of carrier saturation velocity ($\sim$100~\SI{}{\micro\meter}/ns) it can be seen that  the signal amplitude will be significantly reduced already at modest fluence 
if the Ljubljana model is considered due to severe trapping of electrons. CMS model predicts  trapping times twice as large as Atlas model, whose predictions are close to Ljubljana model. 
The trapping times predicted by the Mandi\'c model are bracketed by those of CMS and Atlas.

\section{Validation of Simulations}
\label{sec:validation}

To validate the simulation setup simulation results were compared to data obtained from irradiated samples tested on beam as reported by the CMS tracker 
group~\cite{CMS_passive_CMOS_unirradiated,CMS_irradiated_Pisa_passive_cmos}.  As anticipated, the group realised hybrid planar pixel modules by bonding n-on-p sensors 
realised in a 150~nm CMOS process to a RD53A readout chip. The samples were tested before and after irradiation; results on cluster charge distribution will be discussed in the following.
The cluster charge distribution has been fitted with a Landau function convoluted with a Gaussian; the same choice was made for simulated events, both for unirradiated and irradiated devices.  
The fitted most probable value (MPV) 
was reported as a function of the bias voltage by The CMS tracker group. 

All simulations were carried out using the setup described in Section~\ref{sec:setup}, unless different choices are explicitly mentioned. 
In the following the validation of simulations on unirradiated and irradiated devices will be discussed.

\subsection{Unirradiated devices}
\label{sec:unirrdev}
\begin{figure}[!htb]
\centering
\includegraphics[width=0.66\textwidth]{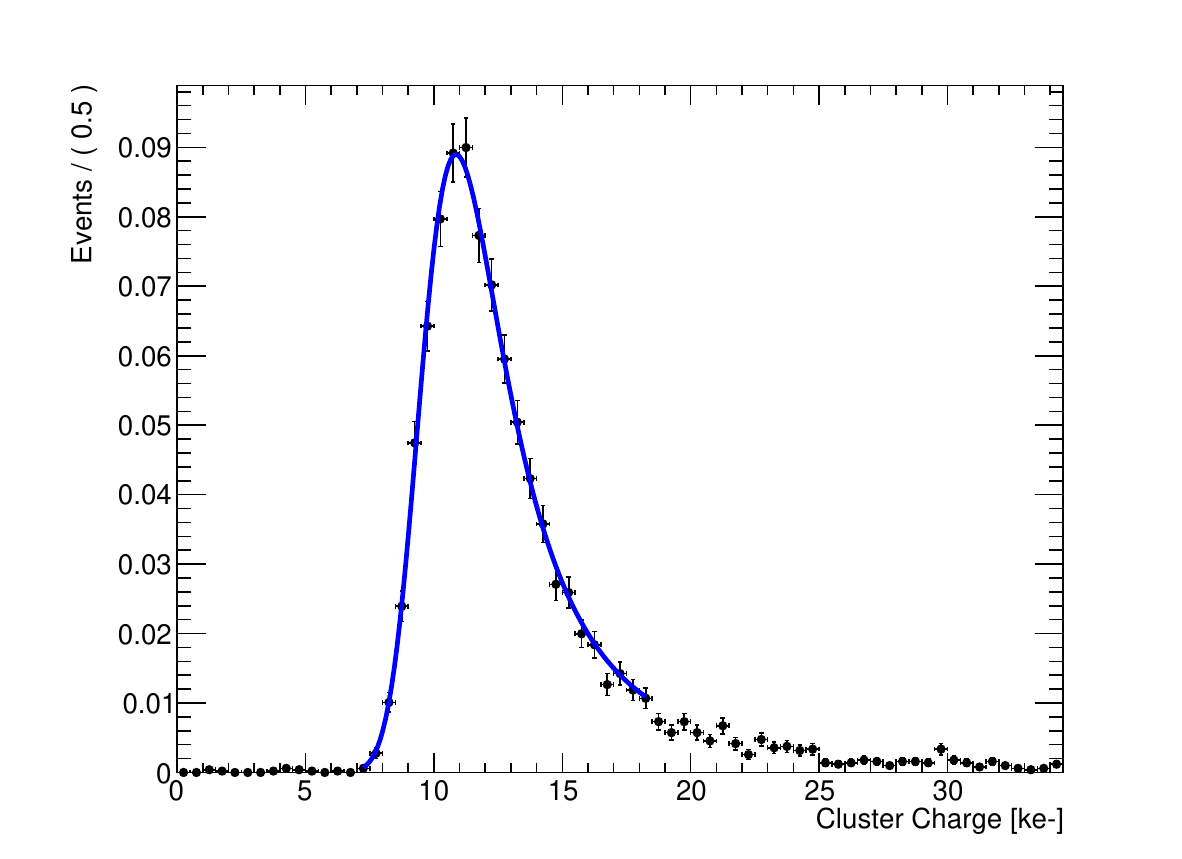}
\caption{\label{fig:cluster_chage_fit_fl0_50V}Cluster charge distribution from simulated events. The sensor was simulated before irradiation and at a bias voltage of 50~V. Points are 
simulated events and the blue line is the result of the fit.}
\end{figure}

To define a proper normalisation for the collected charge an unirradiated device was simulated and results compared to data. 
Both TCAD and Allpix$^2$ simulations were carried out at 20$^{\circ}$~C. 
An example of the fitted cluster charge distribution is shown in Figure~\ref{fig:cluster_chage_fit_fl0_50V} for simulated events at zero fluence and 50~V bias voltage.

In simulation the MPV of cluster charge is calculated as the average of MPV of the Landau function after the fit and the MPV of the fitted convoluted function; the semi-difference of the two is 
an estimate of the fit uncertainty. 

CMS tracker group reported an MPV of the cluster charge of (10.8~$\pm$~0.6)~ke at 50~V; 
the result from simulation is of  (10.91$\pm$0.13)~ke. The agreement between data and simulations is better than 1\%. For this reason in all the rest of the paper simulation results on collected charge will 
be presented and not rescaled to unirradiated values, and compared directly to results from data.

\subsection{Irradiated devices}
\label{sec:irrdev}

The validation of the simulations of irradiated devices has been carried out comparing the results to the performance  of a device irradiated at 
$\Phi = 2.1\times10^{15}$ ~\SI{}{n_{eq}/cm^2} reported by the CMS tracker group. A hit threshold of 1.24~ke was used in both data and simulations.
All four trapping models presented in Section~\ref{sec:trapping} were considered in simulations. 

\begin{figure}[!htb]
\centering
\includegraphics[width=0.66\textwidth]{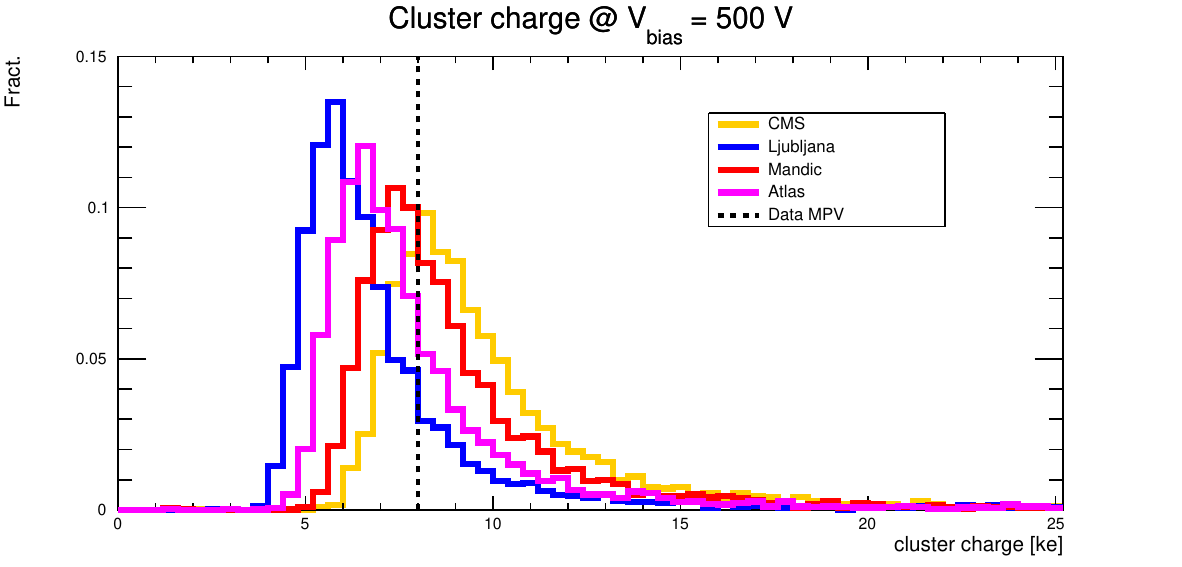}
\caption{\label{fig:cluster_charge_distribution_500V}Cluster charge distribution  from simulated events from a device after a fluence of $\Phi = 2.1\times10^{15}$ ~\SI{}{n_{eq}/cm^2} at 500~V bias voltage. 
Four different trapping models are compared using 
simulated events. The vertical dashed line indicates the MPV of the cluster charge distribution in data.}
\end{figure}

In Figure~\ref{fig:cluster_charge_distribution_500V} results from simulated events after a fluence $\Phi = 2.1\times10^{15}$~\SI{}{n_{eq}/cm^2} are reported when the device was 
at 500~V bias voltage; the MPV of the cluster charge in data at the same conditions is indicated too. 

It is clear from the Figure that the CMS and Mandi\'c trapping models produce the more accurate predictions, with the former slightly overestimating the amount of charge while the latter 
predicting less charge than in data.

\begin{figure}[!htb]
\centering
\includegraphics[width=0.66\textwidth]{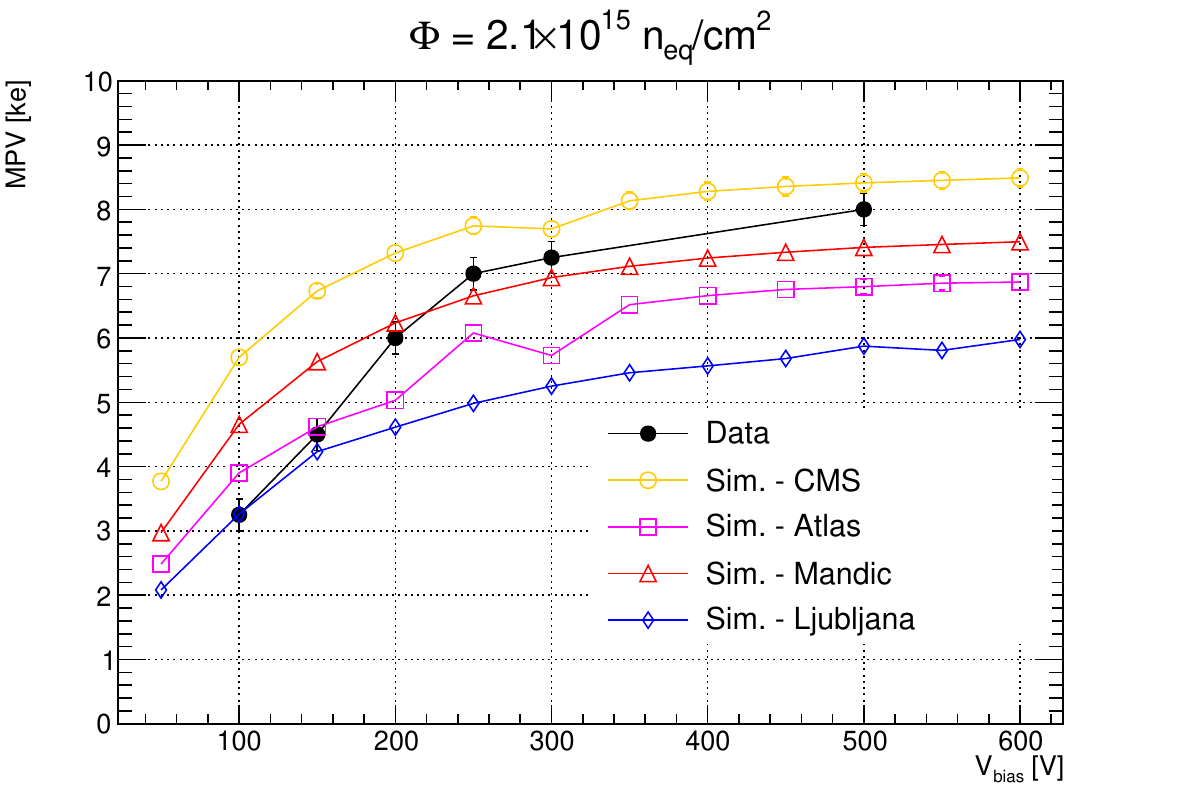}
\caption{\label{fig:trapping_all}MPV of the cluster charge distribution as a function of the bias voltage from data and simulated events for a device after a fluence of $\Phi = 2.1\times10^{15}$ ~\SI{}{n_{eq}/cm^2}. 
Four different trapping models are compared using simulated events.}
\end{figure}

Results on the MPV of cluster charge from  data and simulated events are presented in 
Figure~\ref{fig:trapping_all}. Simulated bias voltages ranged from 50 to 600~V in steps of 50~V.

Two bias voltage ranges can be distinguished, both in data and simulations: a turn-on region (``depletion''), 
where MPV increases rapidly with voltage, ending at around 250~V from where (``depleted'') the increase of charge is weaker 
with increasing voltage. For example in data the MPV increases from about 3 to 7~ke going from 100 to 250~V but then it is limited to 8~ke at 500~V.
This trend is reproduced by all trapping models: this is not surprising as the predictions on trapping times are  of the same order of magnitude (see Table~\ref{tab:trappingtimes}). The good agreement 
about the increase of charge with voltage between data and simulations is the result of the correct modelling of the electric field in TCAD. Hence the choice of the LHCb radiation damage model is validated.
 The correct prediction of the switch between ``depletion'' and ``depleted'' region is extremely important in view of the operations of the ATLAS ITk pixel detector since simulated events 
 can be used to predict the minimal bias voltage for optimal data taking.

Concerning the trapping models, from the same Figure it is clear that the Ljubljana model underestimates the collected charge at almost all voltages. Slight better agreement is obtained by the Atlas model, 
in particular in the ``depletion'' range. The CMS model overestimates the collected charge in all regions, but with a better agreement in the ``depleted'' one. The model whose predictions are closer to data 
is the Mandi\'c one, whose predictions overestimates the charge in the ``depletion'' region but then underestimates it in the ``depleted'' one. A $\chi^2$ test  on the agreement of CMS and Mandi\'c model 
with data indicates the overall better performance of the latter. 
It is interesting to notice that  the Mandi\'c and CMS trapping models essentially bracket the data in the ``depleted'' region. 

\begin{figure}[!htb]
\centering
\includegraphics[width=0.66\textwidth]{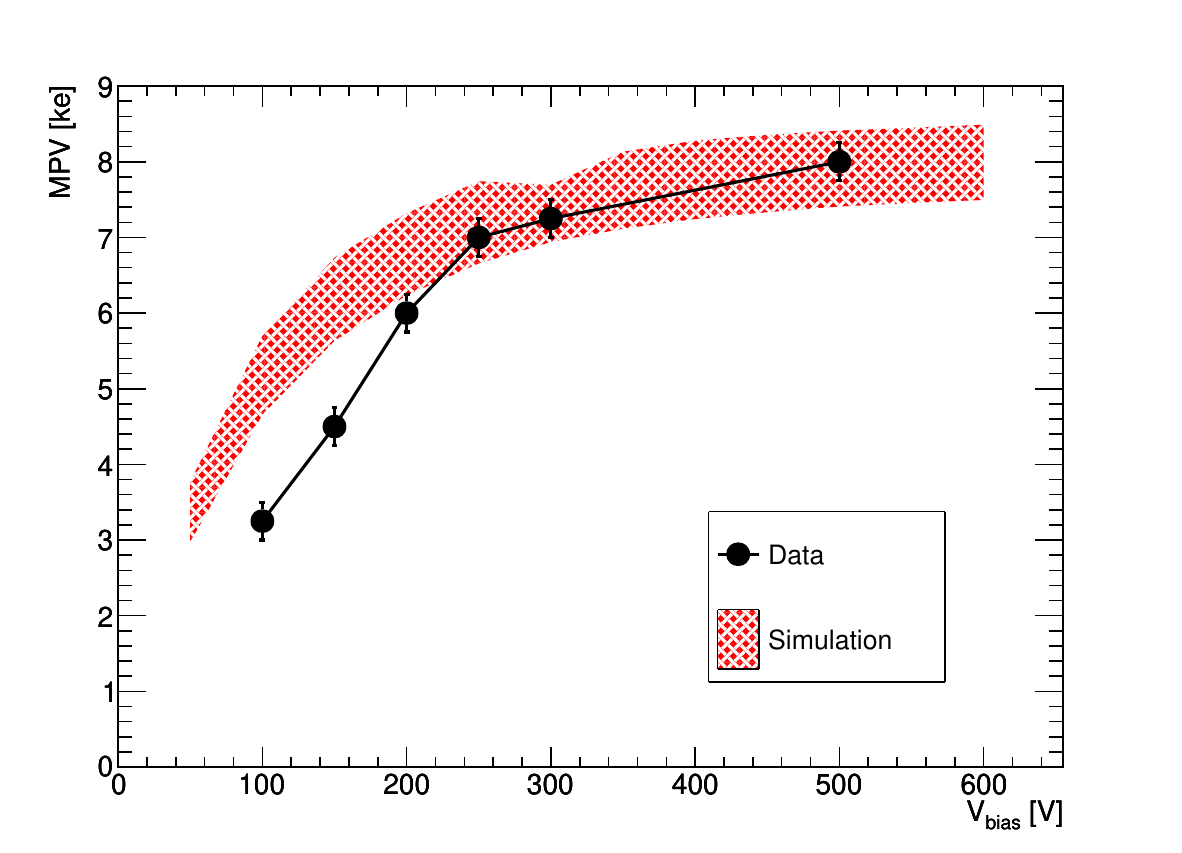}
\caption{\label{fig:data_vs_CMS_Mandic_average}MPV of the cluster charge distribution as a function of the bias voltage from data and simulated events for a device after a fluence of $\Phi = 2.1\times10^{15}$ ~\SI{}{n_{eq}/cm^2}. The simulation predictions are the average of CMS and Mandi\'c ones where the uncertainty is their semi-difference.}
\end{figure}

Given the results reported in Figure~\ref{fig:trapping_all} it seems difficult to choose a single model between Mandi\'c and CMS. If one considers the average of the two predictions  and assigns as uncertainty 
to that the semi-difference the result is as in Figure~\ref{fig:data_vs_CMS_Mandic_average}.

This ``combination'' of predictions correctly covers the experimental results in the ``depletion'' region and provides also some reliable uncertainty estimation. 
As discussed above the Mandi\'c predictions are closer to data than the CMS one but since the latter tend to underestimate systematically the charge in data the ``combination'' is retained as trapping model.  

It is important to stress the excellent agreement between the ``combination'' and the data in the ``depleted'' region: this is the range of voltages to be used by detectors in data taking since it assures 
large and stable values of collected charge which are fundamental for detecting hits with the highest possible efficiency. 

\section{Expected Charge Collection Performance}
\label{sec:performance}

The planar modules of the ATLAS ITk Pixels detector will be integrating fluences up to about $\Phi = 3.5\times10^{15}$ ~\SI{}{n_{eq}/cm^2} by the end of their lifetime. Inner sections of the ITk Pixel detector 
will use n-on-p \SI{100}{\micro\meter} thick planar sensors (``thin'') while the thickness will be of  \SI{150}{\micro\meter} (``thick'') everywhere else. 
In total there will four barrel layers and three endcap ones equipped with planar sensors, all of 50$\times$50~\SI{}{\micro\meter}$^2$ pitch~\cite{itk_tracking_perf_2024}. 

For the ``thin'' sensors fluences of $\Phi =$ 2.5, 3 and 3.5~$\times10^{15}$ ~\SI{}{n_{eq}/cm^2} have been simulated while for ``thick'' ones fluences from 
$\Phi =$ 1.5 to 3.5~$\times10^{15}$ in steps of 0.5~$\times10^{15}$~\SI{}{n_{eq}/cm^2} have been considered; this set of fluences cover all the expected lifetime values for the ITk Pixel detector equipped 
with planar sensors.
The simulation setup is the one reported before. 

To get a sense of the expected hit efficiency a sort of ``effective threshold'' can be calculated as the sum in quadrature of the digitization threshold with 5 times the sum in quadrature of the 
noise and threshold dispersion (see Table~\ref{tab:configuration}); an event with a collected charge above this ``effective threshold'' will certainly be recorded. The value of ``effective threshold'' in this 
simulation setup is of about 1.3~ke. 

\begin{figure}[!htb]
\centering
\includegraphics[width=0.66\textwidth]{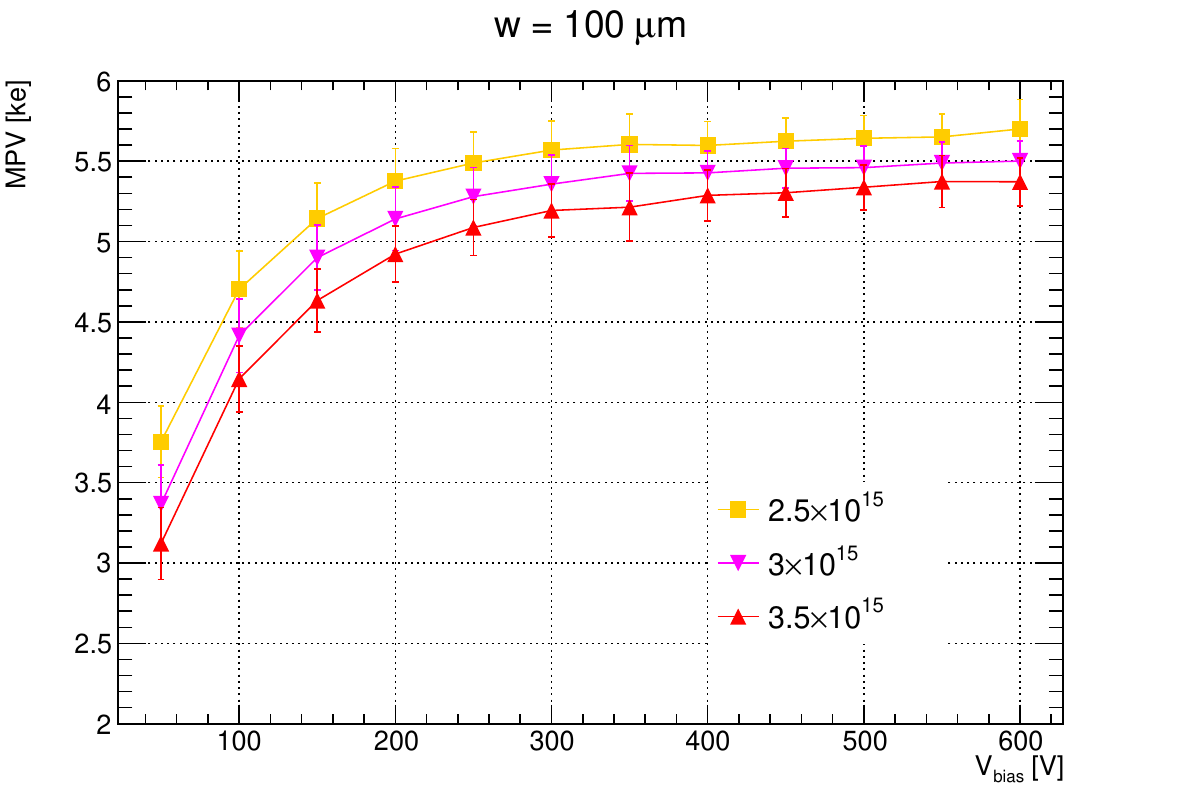}
\caption{\label{fig:w100_simulations_average.pdf}MPV of simulated cluster charge distribution as a function of the bias voltage  for a ``thin'' device after several fluences of irradiation.  The simulation predictions 
are the average of CMS and Mandi\'c ones where the uncertainty is their semi-difference.}
\end{figure}

\begin{figure}[!htb]
\centering
\includegraphics[width=0.66\textwidth]{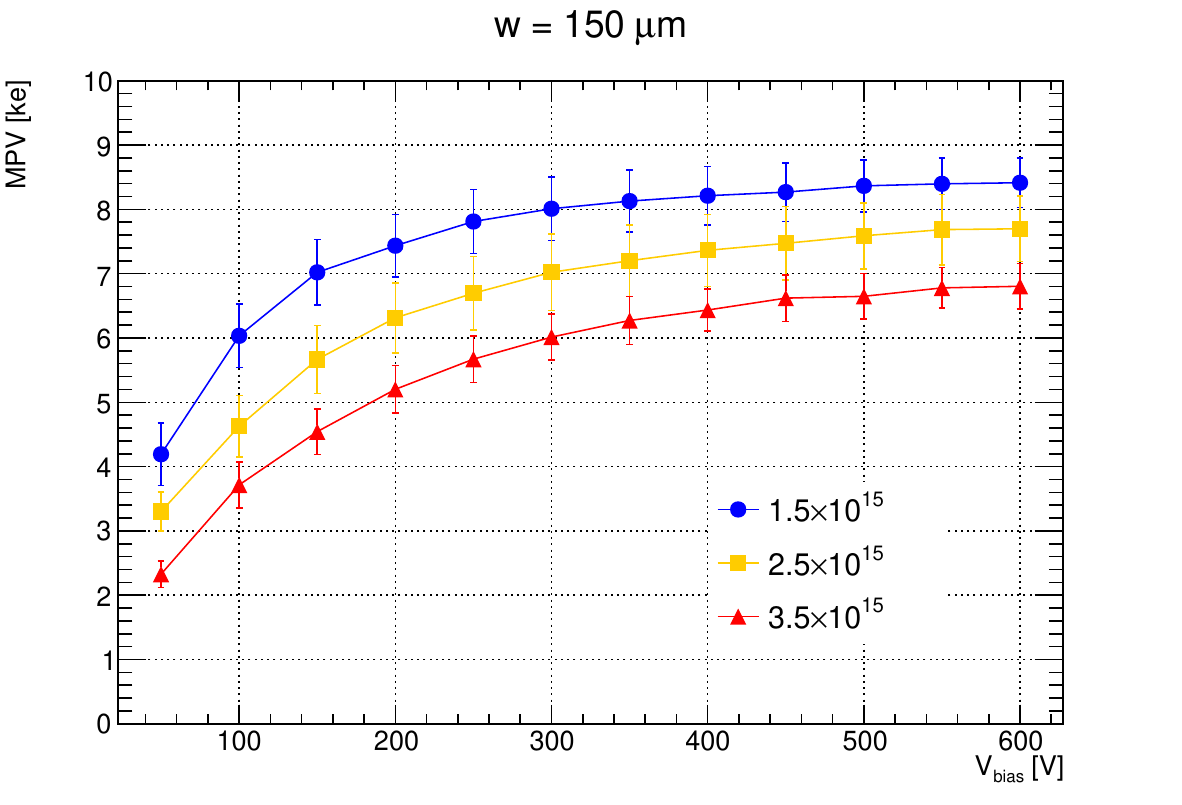}
\caption{\label{fig:w150_simulations_average.pdf}MPV of simulated cluster charge distribution as a function of the bias voltage  for a ``thick'' device after several fluences of irradiation.  The simulation predictions 
are the average of CMS and Mandi\'c ones where the uncertainty is their semi-difference.}
\end{figure}

In Figure~\ref{fig:w100_simulations_average.pdf} and~\ref{fig:w150_simulations_average.pdf} the results of simulations of ``thin'' and ``thick'' samples, respectively; for ``thick'' devices the number of 
results presented is a subset of the total one: this choice has been done to improve the readability of the figure. 

At all fluences and voltages and for both thicknesses enough charge is collected to have full efficiency; this is due also to particles impinging at normal incidence hence giving rise to almost no charge sharing.
For ``thin'' devices the value of collected charge seems to saturate at 300~V while for ``thick'' ones it could be beneficial to run at around 400~V. Nonetheless it seems that it will not be 
necessary to run at the largest possible voltage (600~V for ITk Pixel detector).  
``Thin'' sensors expected to collect in excess of 5~ke at 400~V event at the largest fluence; for ``thick'' sensors the expectation is always above 6~ke at the same voltage. 
 
\begin{figure}[!htb]
\centering
\includegraphics[width=0.49\textwidth]{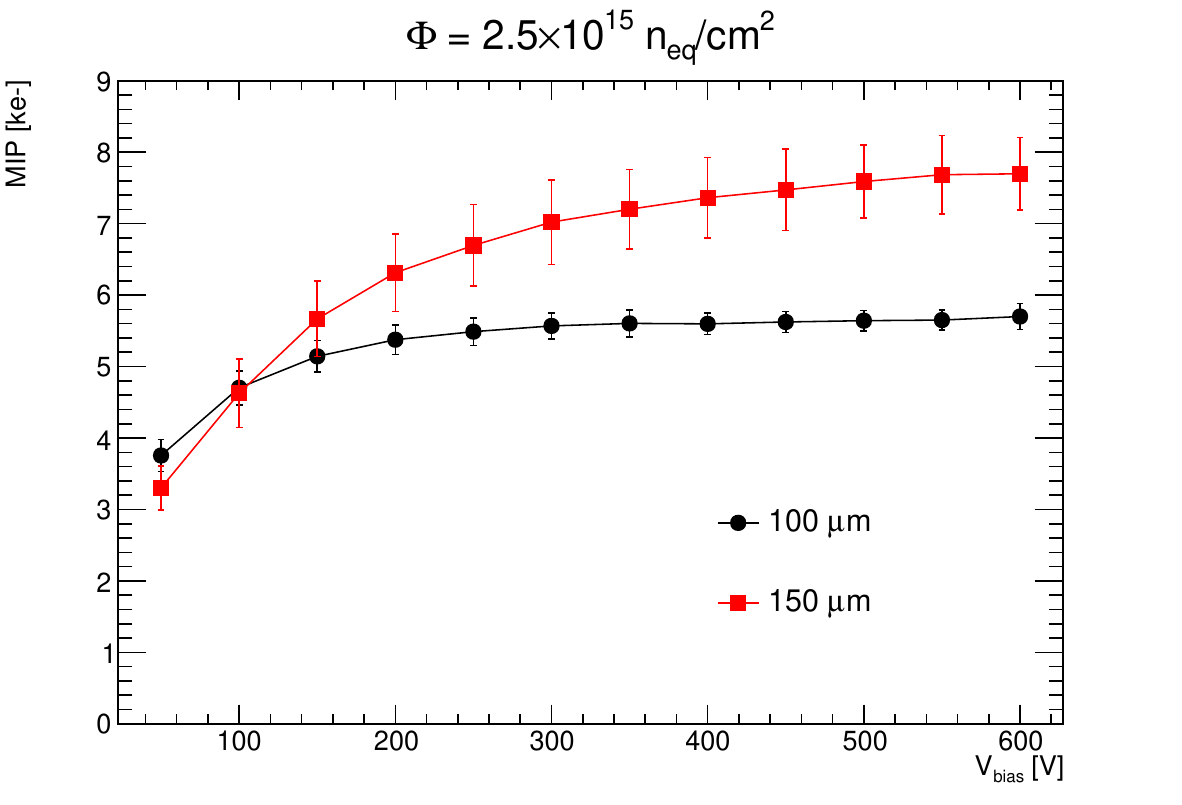}
\includegraphics[width=0.49\textwidth]{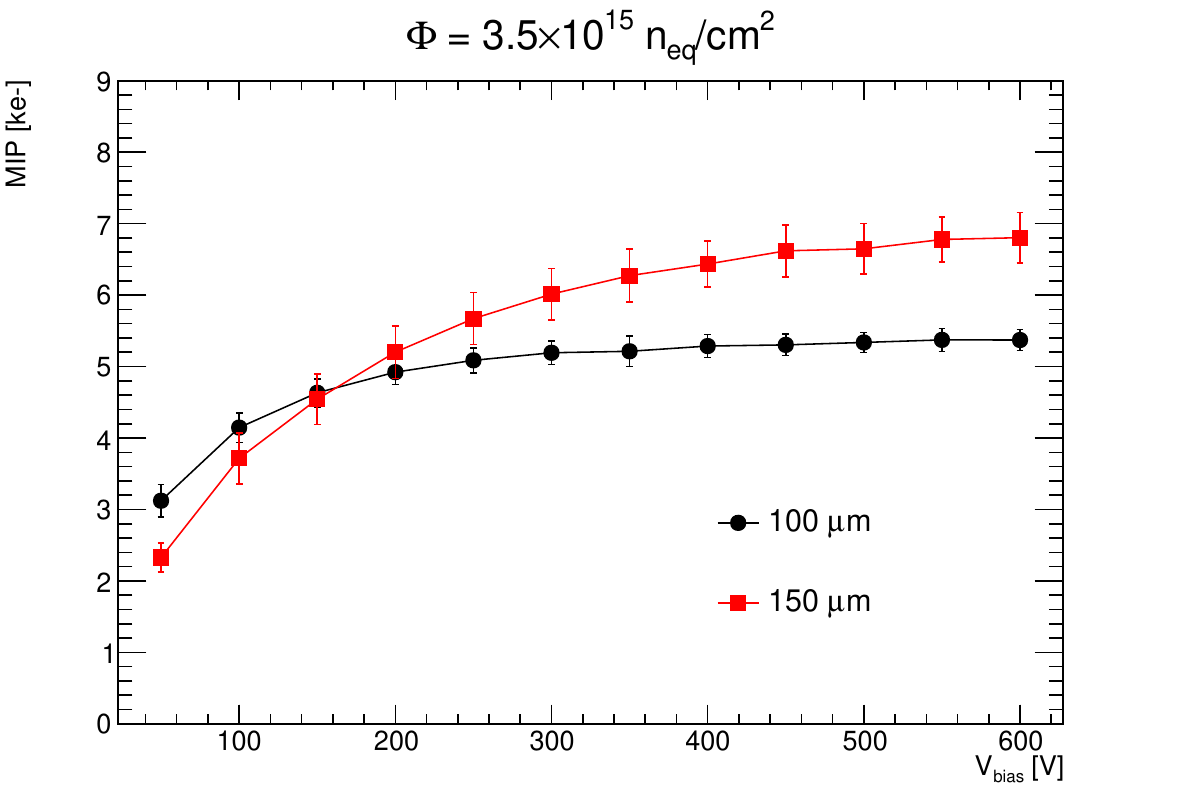}
\caption{\label{fig:average_charge_vs_bias}MPV of simulated cluster charge distribution as a function of the bias voltage after irradiation and different thicknesses.  The simulation predictions 
are the average of CMS and Mandi\'c ones where the uncertainty is their semi-difference. (left)  $\Phi = 2.5\times10^{15}$ ~\SI{}{n_{eq}/cm^2}; (right)  $\Phi = 3.5\times10^{15}$ ~\SI{}{n_{eq}/cm^2}}
\end{figure}

 In Figure~\ref{fig:average_charge_vs_bias} a comparison of performance between the two sensor thicknesses is presented for two fluence values,  $\Phi = 2.5$ and   
 $3.5\times10^{15}$ ~\SI{}{n_{eq}/cm^2}
 
It is interesting to notice that for both fluences,
 apart at very low voltage, the thicker sensors collect more charge than the thinner ones. This is due to the fact that the average free path of carriers is still larger than the 
sensor thickness. Of course a thicker sensor draws more leakage current so it is not advisable to use ``thick'' sensors everywhere given the power budget constraints~\cite{ATLASITkPixelTDR}. 

\begin{figure}[!htb]
\centering
\includegraphics[width=0.66\textwidth]{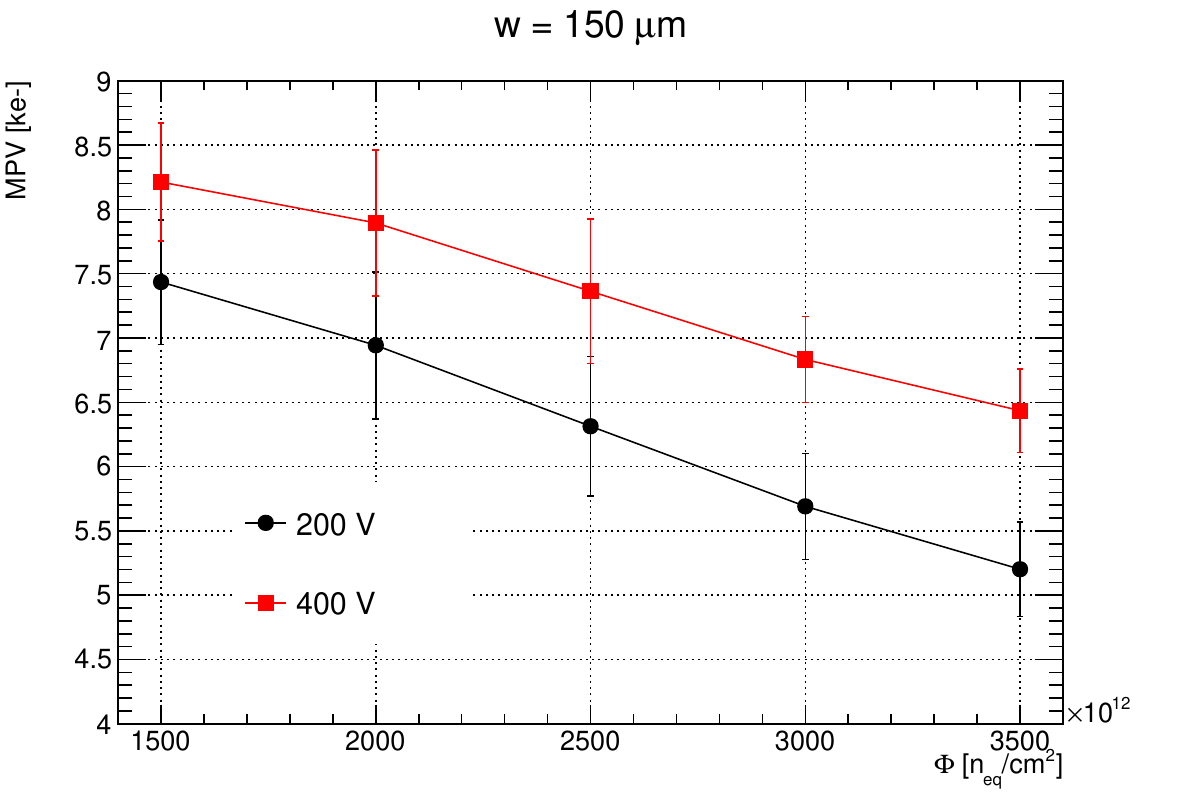}
\caption{\label{figures/w150_average_charge_vs_fluence}MPV of simulated cluster charge distribution as a function of  irradiation fluence for two values of bias voltage. The thickness of the 
sensors was \SI{150}{\micro\meter}.}
\end{figure}

In Figure~\ref{figures/w150_average_charge_vs_fluence} the results for a ``thick'' sensor at 200~V and 400~V are presented as a function of the different irradiation fluences.

This Figure shows the loss of signal amplitude of a pixel module during data taking period.
These results confirm that it should be possible to run at moderate voltage even at the highest fluence.

\section{Conclusions and Outlook}
\label{sec:conclusions}

Hybrid pixel detectors will be exposed to unprecedented fluences at the HL-LHC. Charge trapping will be the limiting factor of performance and it is important to have reliable predictions to make sure 
the tracking and vertexing algorithms can deliver the most precise measurements even when the collected charge is reduced by 40\% or more. 
In this report it was shown that a combination of TCAD and Allpix$^2$ simulations deliver predictions on collected charge that match quite well the value of saturated charge extracted from data. 
It was also shown that the voltage at which the saturation starts can be correctly reproduced.
 
Using this validated combined model predictions on collected charge at expected fluences for the future ATLAS Inner Tracker were prepared. It was shown that the collected charge should 
always be well above the threshold, assuring a large signal-to-threshold ratio. The results also indicate that it will not be necessary to run at the maximal voltage but a few hundreds of volts less should 
anyhow assure full efficiency. 

Similar studies are in preparation for 3D sensors which will be used in the innermost layer of the ATLAS Inner Tracker; these sensors are expected to integrate fluences as high as 
$\Phi = 1.6\times10^{16}$ ~\SI{}{n_{eq}/cm^2} which is almost a factor of 3 larger of what tested in this report. Given the different sensor geometry and fluence range a dedicated study is needed.

The validated simulation model will be used to prepare look-up tables which will be input to the new algorithm~\cite{s24123976} to include radiation damage effects in the ATLAS Monte Carlo event generator. 
It will be then possible to complete the assessment of the robustness of tracking and vertexing algorithms for ITk~\cite{itk_tracking_perf_2024} exploring conditions close to those expected during data taking.

%\appendix
\section*{Acknowledgements}

The author wants to thank Anna Macchiolo of Universit\"at Z\"urich for sharing the CMS tracker group papers on passive CMOS pixels productions. The author is thankful to the Allpix$^2$ developers 
and to Igor Mandi\'c of Jo\v{z}ef Stefan Institute of Ljubljana for the useful suggestions and discussions.

% \end{linenumbers}

%  \bibliographystyle{atlasnote}

  \bibliographystyle{elsarticle-num} 
 \bibliography{biblio.bib}



% \bibliographystyle{JHEP}
 % \bibliography{biblio}

\end{document}